\documentclass[twocolumn,epjc3]{svjour3}

\RequirePackage[T1]{fontenc}

\smartqed  

\RequirePackage{graphicx}
\RequirePackage{mathptmx}      
\RequirePackage{flushend}
\RequirePackage[numbers,sort&compress]{natbib}
\RequirePackage[dvipdfmx, bookmarksnumbered, pdfstartview=FitH,colorlinks,citecolor=blue,urlcolor=blue,linkcolor=blue]{hyperref}
\RequirePackage{amsmath}
\RequirePackage{overpic,graphicx}
\RequirePackage{dcolumn}
\RequirePackage{bm}
\RequirePackage{rotating}
\RequirePackage{subfigure}
\RequirePackage{color}
\RequirePackage{lineno}
\RequirePackage{multirow}

\usepackage{mathptmx}
\DeclareMathAlphabet{\mathcal}{OMS}{cmsy}{m}{n}
\DeclareSymbolFont{largesymbols}{OMX}{cmex}{m}{n}

\journalname{Eur. Phys. J. C}

\begin{document}
\begin{sloppypar}

\title{Feasibility Study of $D_s^+ \to \tau^+ \nu_{\tau}$ Decay and Test of Lepton Flavor Universality with Leptonic $D_s^+$ Decays at STCF}


\author{Huijing Li\thanksref{addr1, addr2, e1}
        \and
        Tao Luo\thanksref{addr3, addr4, e2}
        \and
        Xiaodong Shi\thanksref{addr5, addr6}
        \and
        Xiaorong Zhou\thanksref{addr5, addr6, e3}
}

\thankstext{e1}{email: {lihuijing@htu.edu.cn}
(corresponding author)}
\thankstext{e2}{email: {luot@fudan.edu.cn}
(corresponding author)}
\thankstext{e3}{email: {zxrong@ustc.edu.cn}
(corresponding author)}

\institute{Henan Normal University, Xinxiang 453007, People's Republic of China
\label{addr1}
          \and
          National Demonstration Center for Experimental Physics Education (Henan Normal University), Xinxiang 453007, People's Republic of China\label{addr2}
          \and
         Fudan University, Shanghai 200443, People's Republic of China\label{addr3}
         \and
         Key Laboratory of Nuclear Physics on Ion-Beam Application (MOE) and Institute of Modern Physics, Fudan University, Shanghai 200443, People's Republic of China\label{addr4}
         \and
        University of Science and Technology of China, Hefei 230026, People's Republic of China\label{addr5}
         \and
         State Key Laboratory of Particle Detection and Electronics, Hefei 230026, People's Republic of China\label{addr6}
}

\date{Received: date / Accepted: date}

\maketitle

%

\newcommand{\dstoksk}{D_{s}^{-} \to K_{S}^{0}K^{-}}
\newcommand{\dstokkpi}{D^{-}_{s} \to K^{+}K^{-}\pi^{-}}
\newcommand{\dstokkpipiz}{D^{-}_{s} \to K^{+}K^{-}\pi^{-}\pi^{0}}
\newcommand{\dstokskpipi}{D^{-}_{s} \to K_{S}^{0}K^{-}\pi^{+}\pi^{-}}
\newcommand{\dstokskpipim}{D^{-}_{s} \to K_{S}^{0}K^{+}\pi^{-}\pi^{-}}
\newcommand{\dstopipipi}{D^{-}_{s} \to \pi^{+}\pi^{-}\pi^{-}}
\newcommand{\dstopieta}{D^{-}_{s} \to \pi^{-}\eta_{\gamma \gamma}}
\newcommand{\dstopipizeta}{D^{-}_{s} \to \pi^{-}\pi^{0}\eta_{\gamma \gamma}}
\newcommand{\dstopietapgam}{D^{-}_{s} \to \pi^{-}\eta'_{\pi^{+}\pi^{-}\eta}}
\newcommand{\dstopietaprho}{D^{-}_{s} \to \pi^{-}\eta'_{\gamma \rho^{0}}}
\newcommand{\dstokpipi}{D^{-}_{s} \to K^{-}\pi^{+}\pi^{-}}

\newcommand{\ksk}{K_{S}^{0}K^{-}}
\newcommand{\kkpi}{K^{+}K^{-}\pi^{-}}
\newcommand{\kkpipiz}{K^{+}K^{-}\pi^{-}\pi^{0}}
\newcommand{\kskpipi}{K_{S}^{0}K^{-}\pi^{+}\pi^{-}}
\newcommand{\kskpipim}{K_{S}^{0}K^{+}\pi^{-}\pi^{-}}
\newcommand{\pipipi}{\pi^{+}\pi^{-}\pi^{-}}
\newcommand{\pieta}{\pi^{-}\eta}
\newcommand{\pipizeta}{\pi^{-}\pi^{0}\eta}
\newcommand{\pietapgam}{\pi^{-}\eta'_{\pi^{+}\pi^{-}\eta}}
\newcommand{\pietaprho}{\pi^{-}\eta'_{\gamma \rho^{0}}}
\newcommand{\kpipi}{K^{-}\pi^{+}\pi^{-}}

\newcommand{\etot}{E_{\mathrm{extra}}^{\mathrm{tot}}}

\begin{abstract}
We report a sensitive study of $D_s^+ \to \tau^+ \nu_{\tau}$ decay via $\tau^+ \to e^+ \nu_e \bar{\nu}_{\tau}$ with an integrated luminosity of 1 ab$^{-1}$ at the center-of-mass energy of $4.009$ GeV at a future Super Tau Charm Facility (STCF). 
Under the help of the fast simulation software package, the statistical sensitivity for the absolute branching fraction of $D_s^+ \to \tau^+ \nu_{\tau}$ is determined to be $2\times10^{-4}$. 
Combining with our previous prospect of $D_s^+ \to \mu^+ \nu_{\mu}$, the ratio of the branching fractions for $D_s^+ \to \tau^+ \nu_{\tau}$ over $D_s^+ \to \mu^+ \nu_{\mu}$ can achieve a relative statistical precision of 0.5\%, 
which will provide the most stringent test of the $\tau$-$\mu$ lepton flavor universality in heavy quark decays. 
Taking the decay constant $f_{D_s^+}$ from lattice QCD calculations or the CKM matrix element $|V_{cs}|$ from the CKMfitter group as an input, the relative statistical uncertainties for $|V_{cs}|$ and $f_{D_s^+}$ are estimated to be 0.3\% and 0.2\%, respectively. 


\end{abstract}


\maketitle

\oddsidemargin  -0.2cm
\evensidemargin -0.2cm

\section{\boldmath Introduction}

In the standard model (SM), ignoring radiative corrections, the decay width of $D_s^+ \to \ell^+ \nu_{\ell}$ ($\ell~=~e,~\mu,~\tau$)~\cite{decayratedstotaunu} can be simply given with the formula
\begin{equation}
  \label{decayratedstolnu_draft}
  \Gamma_{D_{s}^{+} \to \ell^+ \nu_\ell} =\frac{G_{F}^{2} f^{2}_{D_s^+} }{8\pi} |V_{cs}|^{2} m^{2}_{\ell^{+}} m_{D_s^+} \left( 1 - \frac{m^{2}_{\ell^{+}}}{m_{D_s^+}^{2}} \right)^{2},
\end{equation}
\noindent where $G_{F}$ is the Fermi coupling constant, $f_{D_s^+}$ is the decay constant parameterizing the strong-interaction physics at the quark-annihilation vertex, $|V_{cs}|$ is the $c \to s$ Cabibbo-Kobayashi-Maskawa (CKM) matrix element, and $m_{\ell^+}$ and $m_{D_s^+}$ are the masses of
the $\ell^+$ lepton and $D_s^+$~\cite{pdg}, respectively. Here and throughout this Letter, charge-conjugate channels are also implied. 
Combining the relationship between the branching fraction (BF) of $D_{s}^{+} \to \ell^+ \nu_\ell$ ($\mathcal{B}_{D_{s}^{+} \to \ell^+ \nu_\ell}$) and its decay width,
\begin{equation}
  \label{br_with_decay_width}
 \mathcal{B}_{D_{s}^{+} \to \ell^+ \nu_\ell} = \tau_{D_s^+} \Gamma_{D_{s}^{+} \to \ell^+ \nu_\ell}
\end{equation}
where $\tau_{D_s^+}$ is the lifetime of $D_s^+$~\cite{pdg}, one can obtain that $\mathcal{B}_{D_{s}^{+} \to \ell^+ \nu_\ell}$ is proportional to the product of $f_{D_s^+}^2$ and $|V_{cs}|^2$. 
Therefore, a precision measurement of $\mathcal{B}_{D_{s}^{+} \to \ell^+ \nu_\ell}$ builds an important bridge to determine the $f_{D_s^+}$ with a given $|V_{cs}|$ from the SM global fit~\cite{pdg}, which plays a key role in calibrating lattice quantum chromodynamics (LQCD) calculations of $f_{D_s^+}$. 
Conversely, one can also extract the $|V_{cs}|$ if taking the $f_{D_s^+}$ from LQCD calculations~\cite{fds_flag} as an input, which provides a rigorous constraint on the unitarity of the CKM matrix.

According to Eq.~(\ref{decayratedstolnu_draft}), the lepton flavor universality (LFU) demands that the ratio of the decay widths for $D_s^+ \to \tau^+ \nu_{\tau}$ and $D_s^+ \to \mu^+ \nu_{\mu}$ only relies on the masses of the $\tau^+$ and $\mu^+$ leptons, which is predicted to be $9.75\pm0.01$~\cite{pdg}. 
Combining the recent measurements of $\mathcal{B}_{D_s^+ \to \tau^+ \nu_{\tau}}$ from BESIII experiment~\cite{dstotaunu_hajime, dstotaunu_panxiang, dstotaunu_hjli} and the world average values of $\mathcal{B}_{D_s^+ \to \tau^+ \nu_{\tau}}$ and $\mathcal{B}_{D_s^+ \to \mu^+ \nu_{\mu}}$~\cite{pdg}, a ratio of $9.67\pm0.34$~\cite{shenxy_talk_charm2020} is obtained, which is consistent with the SM expectation of LFU within one standard deviation. 
However, $BABAR$, LHCb, Belle experiments have reported the hints of LFU violations in semileptonic $B$ decays~\cite{flv1, flv2, flv3, flv4, flv5}. Various theoretical models involving new physics beyond the SM have been used to explain these violations, 
such as leptoquark models~\cite{leptoquarkmodel1, leptoquarkmodel2, leptoquarkmodel3, leptoquarkmodel4, leptoquarkmodel5, leptoquarkmodel6}, $Z^{\prime}$ models~\cite{zpmodels1, zpmodels2, zpmodels3} and two-Higgs-doublet models~\cite{twohiggsmodels1, twohiggsmodels2, twohiggsmodels3}. 
Inspired by the $B$-decay anomalies, theorists analyzed the discrepancies between SM predictions and the experimental measurements in the charm sector, 
and argued that the LFU violations may occur in $c\to s$ transitions due to the interference between the SM amplitude involving a $W^+$ boson and the amplitude including a charged Higgs boson in a two-Higgs-doublet model~\cite{lfu_dstotaunu}, or the interactions with scalar operators~\cite{liying}. 
Therefore, it will be of great interest to measure $\mathcal{B}_{D_{s}^{+} \to \tau^+ \nu_\tau}$ at a higher intensity machine, e.g. Super Tau Charm Facility (STCF), which will achieve higher precision both in statistical and systematic uncertainties.  
 

In this Letter, we perform a feasibility study for the measurement of the absolute $\mathcal{B}_{D_s^+ \to \tau^+ \nu_{\tau}}$ via $\tau^+ \to e^+ \nu_e \bar{\nu}_{\tau}$ with an expected integrated luminosity ($\mathcal{L}$) of 1 ab$^{-1}$ at the center-of-mass energy ($\sqrt{s}$) of 4.009 GeV at STCF. 
Compared to the current best experimental result with a dataset of $\mathcal{L}=$ 6.32 fb$^{-1}$ at $\sqrt{s} = 4.178$-4.226 GeV measured by BESIII experiment~\cite{dstotaunu_hjli}, the study of $D_s^+ \to \tau^+ \nu_{\tau}$ decay at $\sqrt{s}=$ 4.009 GeV at STCF has powerful advantages: 
1) $D_s$ mesons are only produced in $e^+ e^- \to D_s^+ D_s^-$ process~\cite{dsds_crosssetcion}, while other processes of $e^+ e^- \to D_s^{*+} D_s^{-}$ and $e^+ e^- \to D_s^{*+} D_s^{*-}$ are not kinematically allowed;
2) Although the cross section of $e^+ e^- \to D_s^+ D_s^-$ at $\sqrt{s}=$ 4.009 GeV is smaller than that of $e^+ e^- \to D_s^{*+} D_s^{-}$ at $\sqrt{s}=$ 4.178-4.226 GeV~\cite{dsds_crosssetcion}, the larger statistics at $\sqrt{s}=$ 4.009 GeV at STCF will assure a higher precision measurement of $\mathcal{B}_{D_s^+ \to \tau^+ \nu_{\tau}}$; 
3) The charm events produced at threshold are extremely clean, since the backgrounds can be effectively suppressed by the beam energy constraint on $D_s$ mesons; 
4) There is no systematic uncertainty from the photon or $\pi^0$ directly from $D_s^{*+}$ decays. 
The outline of this paper is as follows.
In Sect.~\ref{sec_detector}, we introduce the detector concept for STCF and the related Monte Carlo (MC) samples. 
In Sect.~\ref{sec_st}, we elaborate how to obtain $\mathcal{B}_{D_s^+ \to \tau^+ \nu_{\tau}}$.
In Sect.~\ref{sec_opt}, we explore the optimizations of detector response. 
In Sect.~\ref{sec_results}, we present the results and discussions based on the optimizations.
Finally, we summarize this work in Sect.~\ref{sec_summary}.


\section{\boldmath{Detector and MC simulations}}
\label{sec_detector}
The proposed STCF is a symmetric electron-positron beam collider designed to provide $e^+ e^-$ interactions at $\sqrt{s}=$ 2-7 GeV. The peaking luminosity is expected to be over $0.5\times10^{35}$ cm$^{-2}$s$^{-1}$ at $\sqrt{s}=4$ GeV, and the integrated luminosity per year is $\mathcal{L}=$ 1 ab$^{-1}$. Such an environment will be an important low-background playground to test the SM and probe possible new physics beyond the SM. 
The STCF detector is a general purpose detector designed for $e^+ e^-$ collider which includes a tracking system composed of the inner and outer trackers, a particle identification (PID) system with excellent $K/\pi$ separation power, and an electromagnetic calorimeter (EMC) with an excellent energy resolution and a good time resolution, a super-conducting solenoid and a muon detector (MUD) that provides good charged $\pi/\mu$ separation. The detailed conceptual design for each sub-detector, the expected detection efficiency and resolution can be found in Refs.~\cite{penghp_talk, luoq_talk, fast_simu_tool}. 

At present, the STCF detector and the corresponding offline software system are under research and development. 
To access the physics study, a fast simulation tool for STCF has been developed~\cite{fast_simu_tool}, which takes the most common event generators as input to perform a fast and realistic simulation. The simulation includes resolution and efficiency responses for tracking of final state particles, PID system and kinematic fit related variables. By default, all the parameterized parameters for each sub-detector performance are based on the BESIII performance~\cite{Ablikim:2009aa}, but can be adjusted flexibly by scaling a factor according to the expected performance of the STCF detector, which can be used to optimize the detector design according to physical requirements. 


The MC samples are generated based on the STCF fast simulation tool. The first sample is a generic MC sample simulated at $\sqrt{s}=4.009$ GeV with $\mathcal{L}=$ 0.1 ab$^{-1}$. 
It includes open charm processes, continuum light quark production, QED processes, and initial-state radiation (ISR) processes. 
The second sample is an exclusive signal MC sample of $e^+ e^- \to (\gamma_{\rm ISR}) D_s^+ D_s^-$, where one $D_s^+$ decays to $\tau^+ \nu_{\tau}$ with $\tau^+ \to e^+ \nu_e \bar{\nu}_{\tau}$, and the other $D_s^-$ is fully reconstructed. In the simulation, $e^+ e^-$ collisions are simulated by the {\sc kkmc}~\cite{kkmc} generator, which takes into account the beam energy spread and the ISR correction, where the beam energy spread is assigned to the same value as that of BEPCII~\cite{shixd_cp_paper}. The known decay modes are generated with {\sc evtgen}~\cite{evtgen} with BFs set to the world average values~\cite{pdg}, while the unmeasured decays are generated with {\sc lundcharm}~\cite{lundcharm}.

This study is performed with the generic MC sample. The signal MC sample is utilized to obtain the detection efficiency and optimize the STCF detector responses.

\section{\boldmath{Analysis}}
\label{sec_st}
\subsection{\boldmath{Measurement technique}}
Benefiting from the $D_s^+ D_s^-$ pair produced in $e^+ e^-$ collision at $\sqrt{s}=4.009$ GeV, a ``double tag" (DT) technique pioneered by the MARK-III Collaboration~\cite{dtmethod1, dtmethod2} can be employed to measure the absolute $\mathcal{B}_{D_s^+ \to \tau^+ \nu_{\tau}}$. 
We select ``single tag" (ST) events in which a tag $D_s^-$ is fully reconstructed. And then we look for the signal $D_s^+$ decays of interest in the remainder of each event, called as DT events. 
The decay chain is $e^+ e^- \to (\gamma_{\rm ISR}) D_s^{+}D_s^{-}$, where the tag $D_s^-$ is fully reconstructed with 11 tag modes:~$\dstoksk$, $\kkpi$, $\kkpipiz$, $\kskpipi$, $\kskpipim$, $\pipipi$, $\pieta$, $\pipizeta$, $\pietapgam$, $\pietaprho$, $\kpipi$, 
of which the intermediated states are reconstructed by $K_S^0 \to \pi^+ \pi^-$, $\pi^0/\eta\to \gamma \gamma$, $\eta^{\prime}_{\pi^+ \pi^- \eta} \to \pi^+ \pi^- \eta$, $\eta^{\prime}_{\gamma \rho^0} \to \gamma \rho^0$, and $\rho^0 \to \pi^+ \pi^-$. The signal side is $D_s^+\to\tau^+ \nu_{\tau},~\tau^+\to e^+ \nu_e \bar{\nu}_{\tau}$ that only has one charged track identified as $e^+$. 
 
The absolute BF of $D_s^+ \to \tau^+ \nu_{\tau}$ decay for a tag mode $\alpha$ is calculated by
\begin{eqnarray}
  \label{dtbr}
  \mathcal{B}_{ D_s^+ \to \tau^+ \nu_{\tau}}^{\alpha} = \frac{N_{\rm DT}^{{\rm obs},\,\alpha} / \epsilon^{\alpha}_{\rm DT}}{ N_{\rm ST}^{{\rm obs},\,\alpha}/\epsilon_{\rm ST}^{\alpha} \cdot \mathcal{B}_{\tau^+ \to e^+ \nu_e \bar{\nu}_{\tau}}},
\end{eqnarray}
where $N_{\rm ST}^{{\rm obs},\,\alpha}$ and $N_{\rm DT}^{\rm obs,\,\alpha}$ are the ST and DT yields, $\epsilon_{\rm ST}^{\alpha}$ and $\epsilon_{\rm DT}^{\alpha}$ are the ST and DT efficiencies, and $\mathcal{B}_{\tau^ \to e^+ \nu_e \bar{\nu}_{\tau}}$ is the BF of $\tau^+ \to e^+ \nu_e \bar{\nu}_{\tau}$ quoted from the world average value~\cite{pdg}. 
Combined with results from 11 tag modes, the final BF ($\mathcal{B}_{D_s^+ \to \tau^+ \nu_{\tau}}$) and corresponding error ($\partial\mathcal{B}_{D_s^+ \to \tau^+ \nu_{\tau}}$) are subsequently estimated by
\begin{eqnarray}
  \label{errdtbr}
  \mathcal{B}_{ D_s^+ \to \tau^+ \nu_{\tau}} =&&\frac{\sum_{\alpha} \omega^{\alpha} \mathcal{B}^{\alpha}_{ D_s^+ \to \tau^+ \nu_{\tau}}} { \sum_{\alpha} \omega^{\alpha}}, \nonumber\\
  \partial\mathcal{B}_{ D_s^+ \to \tau^+ \nu_{\tau}} = &&\frac{1}{\sqrt{\sum_{\alpha} \omega^{\alpha}}},\nonumber\\
  \omega^{\alpha} =&& \frac{1}{(\partial \mathcal{B}^{\alpha}_{ D_s^+ \to \tau^+ \nu_{\tau}})^2},
\end{eqnarray}
where $\partial \mathcal{B}^{\alpha}_{ D_s^+ \to \tau^+ \nu_{\tau}}$ is the statistical uncertainty of the BF for the tag mode $\alpha$.


\subsection{\boldmath{Single tag}}
\label{sec_st_ana}

Each charged track is demanded to satisfy the vertex requirement and detector acceptance in fast simulation. The combined confidence levels for the pion and kaon hypotheses ($CL_{\pi}$ and $CL_{K}$, respectively) are estimated, and the particle type with the higher confidence level is assigned to each track.

The $\pi^0$ ($\eta$) candidates are reconstructed from pairs of photons. The invariant mass ($M_{\gamma \gamma}$) of two photons is required to be within [0.115, 0.150] GeV/$c^2$ for $\pi^0$ candidates, and [0.50, 0.57] GeV/$c^2$ for $\eta$ candidates. In the following analysis, the photon pair is kinematically constrained to be the nominal mass of the $\pi^0$ ($\eta$) to improve the resolution of $\pi^0$ ($\eta$) momentum. The $K_S^0$ candidates are reconstructed with pairs of oppositely charged tracks that assumed to be pions and further required to have an invariant mass ($M_{\pi^+ \pi^-}$) of $\pi^+ \pi^-$ within [0.487, 0.511] GeV/$c^2$. The $\rho^0$ candidates are required to be within a range of $M_{\pi^+\pi^-} >$ 0.5 GeV/$c^2$. For $\dstokpipi$ tag mode, the $M_{\pi^+\pi^-}$ is outside the range of [0.480, 0.515] GeV/$c^2$ to avoid the overlap with the $\dstoksk$ tag mode. The $\eta^{\prime}$ candidates are demanded to have an invariant mass of $\pi^+ \pi^- \eta$ within [0.946, 0.97] GeV/$c^2$, or an invariant mass of $\gamma \rho^0$ within [0.94, 0.976] GeV/$c^2$. The momenta of charged and neutral pions are larger than 0.1 GeV/$c$ to suppress the soft pions from $D^{*}$ decays, and the momentum of photon from $\eta^{\prime}$ decay is greater than 0.1 GeV/$c$.

To identify the reconstructed ST $D_s^-$ candidates, we use two variables, the beam-constrained mass, $M_{\rm BC}$, and the energy difference, $\Delta E$, which are defined as $M_{\rm BC} \equiv \sqrt{E^2_{\rm beam}-|\vec{p}_{D_s^-}|^2}$, $\Delta E \equiv E_{D_s^-}-E_{\rm beam}$. 
Here, $\vec{p}_{D_s^-}$ and $E_{D_s^-}$ are the reconstructed momentum and energy of the $D_s^-$ candidate in the $e^+ e^-$ center-of-mass system, and $E_{\rm beam}$ is the beam energy. We accept ST $D_s^-$ candidates with $M_{\rm BC}$ greater than 1.90 GeV/$c^2$ and mode-dependent $\Delta E$ requirements of approximately 3 standard deviations, as listed in Table~\ref{deltae_signal_region}. For an event, only one ST $D_s^-$ candidate per mode per charge is accepted with the smallest $|\Delta E|$.

\begin{table}[htbp]
\centering
\caption{$\Delta E$ requirements for 11 tag modes.}
\label{deltae_signal_region}
 \setlength{\extrarowheight}{1.2ex}
 \setlength{\tabcolsep}{7pt}
 \renewcommand{\arraystretch}{1.0}
  \vspace{0.1cm}
\begin{tabular}{p{3.0cm} m{3.0cm}<{\raggedleft}}\hline
Mode & $\Delta E$ (MeV) \\\hline
$\dstoksk$	    &$[	-23.2	,~	22.5	]$\\
$\dstokkpi$	    &$[	-25.6	,~	24.5	]$\\
$\dstokkpipiz$	&$[	-30.0	,~	25.0	]$\\
$\dstokskpipi$	&$[	-28.4	,~	28.0	]$\\
$\dstokskpipim$	&$[	-27.9	,~	27.1	]$\\
$\dstopipipi$	  &$[	-24.4	,~	22.7	]$\\
$\dstopieta$	  &$[	-35.0	,~	35.0	]$\\
$\dstopipizeta$	&$[	-45.0	,~	25.0	]$\\
$\dstopietapgam$&$[	-31.3	,~	30.4	]$\\
$\dstopietaprho$&$[	-34.9	,~	24.7	]$\\
$\dstokpipi$	  &$[	-19.4	,~	18.8	]$\\
\hline
\end{tabular}
 \vspace{-0.1cm}
\end{table}

\begin{figure}[tb]
\centering
   \includegraphics[width=0.48\textwidth]{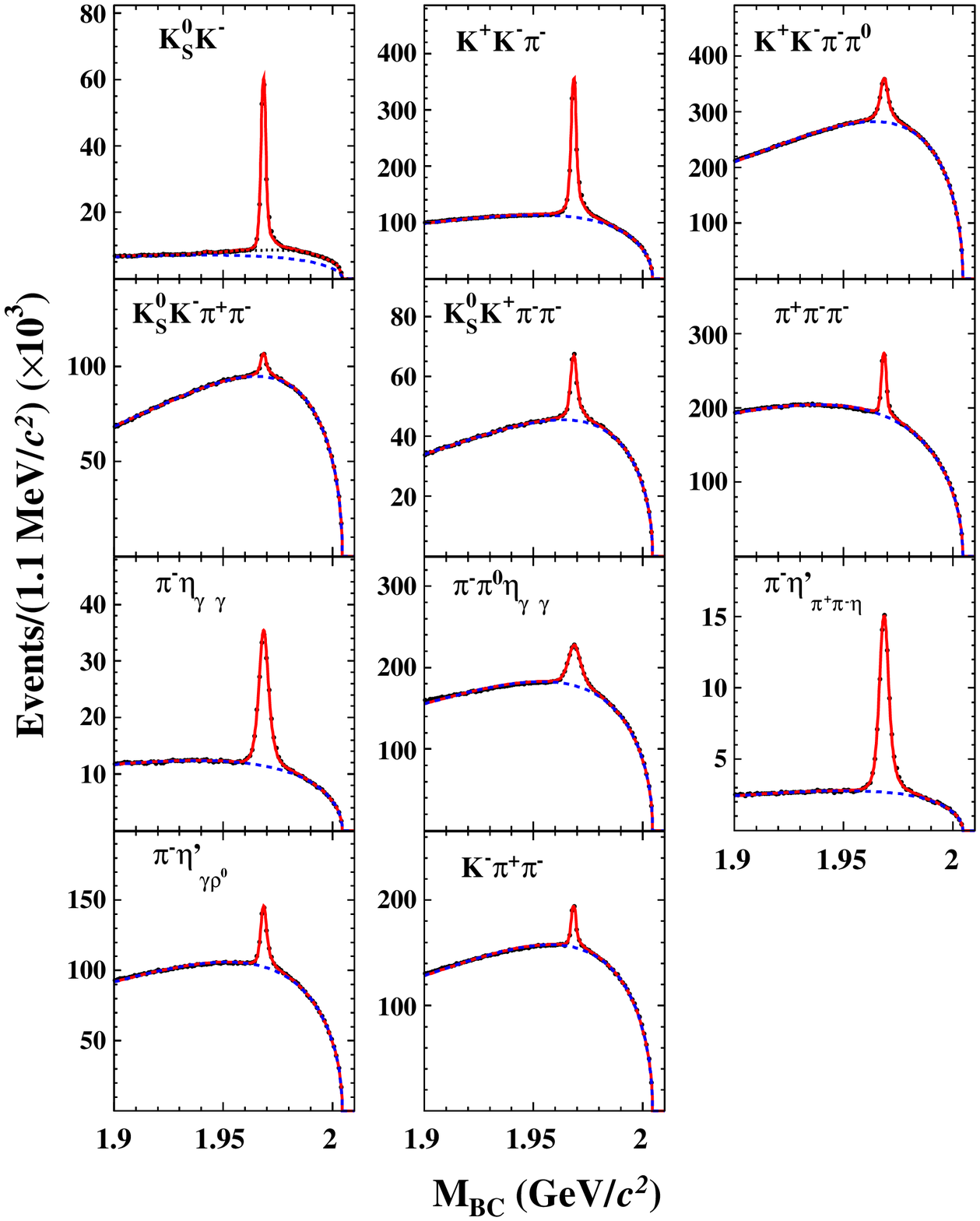}
  \caption{Fitting results of $M_{\rm BC}$ distributions for 11 tag modes. The dots with error bars are from generic MC sample, the solid red lines are the best fits, the dashed blue lines are the background shapes, and the difference between the dashed blue lines and the dotted black line for $\dstoksk$ tag mode is the peaking background $D^-\to K_S^0 \pi^-$.}
\label{ST_yield_data}
\end{figure}

The ST yields are obtained from fitting $M_{\rm BC}$ distributions, as shown in Fig.~\ref{ST_yield_data}. In the fit, the signal shape is modeled by the MC simulation, and the combinatorial background is described by an ARGUS function~\cite{argus}. For the tag mode $\dstoksk$, the shape of the peaking background $D^-\to K_S^0 \pi^-$ is extracted from the MC simulation, and the size is floated. The ST yields are evaluated with the $M_{\rm BC}$ signal region of [1.96, 1.98] GeV/$c^2$. With the same fitting procedure, the ST efficiencies are estimated by fitting to $M_{\rm BC}$ distributions from one of sixth generic MC sample, where the correlations with the total one are assumed to be negligible. Table~\ref{measured_br_dstotaunu} lists the ST yield and ST efficiency for each tag mode.

\begin{table*}[htbp]
\begin{center}
\caption{The ST (DT) yield ($N_{\rm ST(DT)}^{{\rm obs}}$), ST (DT) efficiency ($\epsilon_{\rm ST(DT)}$), signal efficiency ($\epsilon_{\rm sig} = \epsilon_{\rm DT}/\epsilon_{\rm ST}$) for each tag mode, as well as the measured $\mathcal{B}_{D_s^+\to\tau^+\nu_{\tau}}$. ``Average" is the inverse uncertainty weighted BF. The BFs of the intermediated states ($K_S^0$, $\pi^0$, $\eta$, $\eta'$, and $\rho^0$) decays are not included both in the ST and DT efficiencies. All of the uncertainties are statistical only.}
\label{measured_br_dstotaunu}
\setlength{\extrarowheight}{1.2ex}
 \setlength{\tabcolsep}{7pt}
  \renewcommand{\arraystretch}{1.0}
  \vspace{0.1cm}
\begin{tabular}{l  r@{$\,\pm\,$}l  r@{$\,\pm\,$}l  r@{$\,\pm\,$}l  r@{$\,\pm\,$}l  r@{$\,\pm\,$}l r@{$\,\pm\,$}l }\hline
Mode & \multicolumn{2}{c}{$N_{\rm ST}^{\rm obs}$} & \multicolumn{2}{c}{$\epsilon_{\rm ST}~(\%)$}  & \multicolumn{2}{c}{$N_{\rm DT}^{\rm obs}$} & \multicolumn{2}{c}{$\epsilon_{\rm DT}~(\%)$} & \multicolumn{2}{c}{$\epsilon_{\rm sig}~(\%)$} & \multicolumn{2}{c}{$\mathcal{B}_{D_s^+\to\tau^+\nu_{\tau}}~(\%)$} \\\hline
$\dstoksk$	&	160505 	&	493 	&	40.74 	&	0.30 	&	994 	&	49 	&	26.91	&	0.07	&	66.04	&	0.52	&	5.27	&	0.26	\\
$\dstokkpi$	&	904457 	&	1418 	&	42.41 	&	0.16 	&	6036 	&	119 	&	28.60	&	0.07	&	67.44	&	0.31	&	5.57	&	0.11	\\
$\dstokkpipiz$	&	417245 	&	1908 	&	15.80	&	0.17 	&	2804 	&	91 	&	11.83	&	0.05	&	74.85	&	0.88	&	5.05	&	0.16	\\
$\dstokskpipi$	&	53670 	&	978 	&	18.04 	&	0.56 	&	378 	&	51 	&	12.88	&	0.05	&	71.38	&	2.24	&	5.55	&	0.76	\\
$\dstokskpipim$	&	97290 	&	742 	&	20.12 	&	0.37 	&	680 	&	48 	&	13.95	&	0.05	&	69.31	&	1.31	&	5.67	&	0.40	\\
$\dstopipipi$	&	260620 	&	1168 	&	56.09 	&	0.61 	&	1665 	&	101 	&	36.45	&	0.08	&	64.98	&	0.71	&	5.53	&	0.34	\\
$\dstopieta$	&	135058 	&	535 	&	49.30 	&	0.47 	&	909 	&	42 	&	33.95	&	0.07	&	68.87	&	0.68	&	5.50	&	0.26	\\
$\dstopipizeta$	&	351782 	&	1680 	&	25.65 	&	0.30 	&	2650 	&	79 	&	18.74	&	0.06	&	73.07	&	0.88	&	5.80	&	0.17	\\
$\dstopietapgam$	&	68572 	&	331 	&	24.65 	&	0.29 	&	443 	&	29 	&	16.91	&	0.10	&	68.60	&	0.89	&	5.30	&	0.35	\\
$\dstopietaprho$	&	170640 	&	984 	&	33.95 	&	0.49 	&	1173 	&	78 	&	23.44	&	0.09	&	69.04	&	1.04	&	5.60	&	0.37	\\
$\dstokpipi$	&	139359 	&	1123 	&	46.12 	&	0.92 	&	1039 	&	84 	&	33.79	&	0.07	&	73.25	&	1.47	&	5.72	&	0.46	\\\hline
Average	&	\multicolumn{10}{c}{}																			&	5.49	&	0.07	\\
\hline
\end{tabular}
\vspace{-0.1cm}
\end{center}
\end{table*}

%

\subsection{\boldmath{Double tag}}

In the presence of the selected ST $D_s^-$ candidate, we look for $D_s^+\to\tau^+ \nu_{\tau}$ signals at the recoil side against of the ST $D_s^-$, where $\tau^+\to e^+ \nu_e \bar{\nu}_{\tau}$. It is required that there is only one charged track with the opposite charge with the ST $D_s^-$. The charged track is identified as positron with the requirements of $CL_e/[CL_e + CL_\pi + CL_K] >0.8$, the momentum greater than 0.2 GeV/$c$, and the ratio of the deposited energy in the EMC over the momentum larger than 0.8.

\begin{figure*}[htbp]
\centering
   \includegraphics[width=0.9\textwidth]{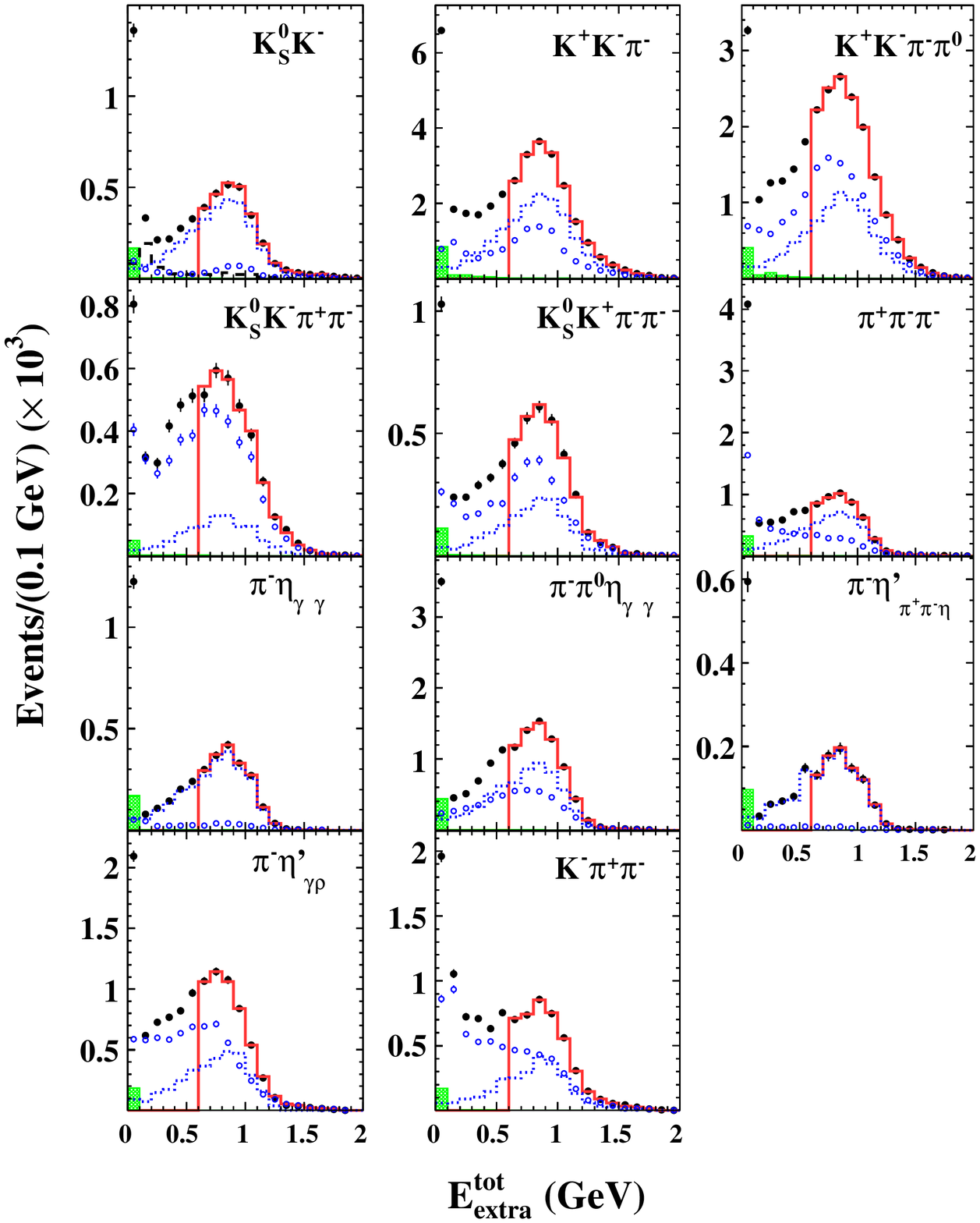}
  \caption{Fitting results of $\etot$ distributions for 11 tag modes. The dots with error bars are from generic MC sample, the solid red lines are the best fits, the open dots with error bars are the non-$D_s^-$ background, the hatched green lines are the peaking background of $D_s^+ \to K_L^0 e^+ \nu_e$, the dotted blue lines are the $D_s^+ \to X e^+ \nu_e$ background, and the dashed black line for $\dstoksk$ tag mode is another peaking background $D^-\to K_S^0 \pi^-$.}
\label{fit_etot_data_4178}
\end{figure*}

The variable $\etot$ is exploited to demonstrate the signal $D_s^+$ candidate, which is defined as the total energy of showers in the EMC, except for those used in the ST side and the photons from positron satisfying that the angle with the positron be less than 5 degrees. Since there is no additional showers, the signal events peak at 0 in $\etot$ distribution, as shown in Fig.~\ref{fit_etot_data_4178}. The signal region of $\etot$ is required to be less than 0.4 GeV/$c$, which is optimized with the MC samples.

The background sources are divided into three categories according to MC studies. The first one is the non-$D_s^-$ background, of which the ST $D_s^-$ is reconstructed incorrectly. The second one is the $D_s^+$ peaking background of $D_s^+ \to K_L^0 e^+ \nu_e$ due to little or no deposited energy in the EMC for the $K_L^0$ meson, while there is another peaking background $D^-\to K_S^0 \pi^-$ for the tag mode $D_s^- \to K_S^0 K^-$ because the bachelor pion is mis-identified as kaon. The last one is the $D_s^+$ non-peaking backgrounds $D_s^+ \to X e^+ \nu_e$, excluding $D_s^+\to\tau^+ \nu_{\tau},~\tau^+\to e^+ \nu_e \bar{\nu}_{\tau}$ and $D_s^+ \to K_L^0 e^+ \nu_e$ decays. The latter two categories are dominantly accompanied with the correctly reconstructed ST $D_s^-$.

In terms of how to obtain the DT yield, the methods related with the signal shape are not adopted, since it is not trivial to accurately model the signal shape of $\etot$. 
On the contrary, we perform the binned maximum likelihood fit to $\etot$ in the high side region of $\etot >$ 0.6 GeV, which are dominant by the backgrounds and the effect from signal tail can be negligible. 
In the fit, the shape of the non-$D_s^-$ background is modelled with the events in $M_{\rm BC}$ sideband regions of [1.915, 1.935] $\cup$ [1.990, 2.000] GeV/$c^2$, and the size is fixed according to the normalized $M_{\rm BC}$ sideband events. For tag modes with neutral daughters, the resolution in $M_{\rm BC}$ is degraded, and we must correct for true signal events that populate the sideband based on MC simulations. For the peaking backgrounds, the shapes and the sizes of $D_s^+ \to K_L^0 e^+ \nu_e$ and $D^-\to K_S^0 \pi^-$ decays are determined from MC simulations.
For the $D_s^-\to X e^+ \nu_e$ background, the shape is also extracted from MC simulation, and the size is floated in the fit. Figure~\ref{fit_etot_data_4178} shows the fitting results of $\etot$. 
After extrapolated the backgrounds into the signal region of $\etot<$ 0.4 GeV, the DT yields are estimated by subtracting all of the background yields within the signal region of $\etot<$ 0.4 GeV from the number of observed events in this region. 
The DT efficiency is obtained from the signal MC sample. Table~\ref{measured_br_dstotaunu} lists the DT yield and DT efficiency for each tag mode.

The measured BF of $D_s^+\to\tau^+ \nu_{\tau}$ for each tag mode is calculated according to Eq.~(\ref{dtbr}), as listed in Table~\ref{measured_br_dstotaunu}. After weighted by the statistical uncertainty of each tag mode as shown in Eq.~(\ref{errdtbr}), the BF of $D_s^+\to\tau^+ \nu_{\tau}$ is evaluated to be $\mathcal{B}_{D_s^+\to\tau^+ \nu_{\tau}}=(5.49\pm0.07_{\rm stat})\times 10^{-2}$, where the uncertainty is the statistical only. This result is consistent with the input value $5.54\times10^{-2}$ when generating the generic MC sample within one standard deviation, which proves that the whole procedure is reasonable.

\section{\boldmath{Optimization of detector response}}
\label{sec_opt}
The loss of the detection efficiency mainly arises from the effects of charged and neutral tracking selections, and the mis-identification rates for $\pi/K$. 
These effects correspond to the sub-detectors of the tracking system, the EMC and the PID system. Since the backgrounds in the DT side are from the events with the incorrectly reconstructed ST $D_s^-$ candidate or the true positron candidate accompanied by other particles, 
we only investigate the ST efficiency and the corresponding figure-of-merit defined by $S/\sqrt{S+B}$ along with the variations of the sub-detector's responses,  and then the requirement of detector design can be optimized accordingly. 
Here $S$ and $B$ denote the expected ST yield and the background yield, respectively.

In the fast simulation~\cite{fast_simu_tool}, by default, all parameters for each sub-detector performance are parameterized based on the BESIII detector~\cite{Ablikim:2009aa} that operates in the similar energy region, but can be adjusted flexibly by a scale factor according to the expected performance of the STCF detector, or by implementing a special interface to model any performance described with an external histogram, an input curve, or a series of discrete data~\cite{fast_simu_tool}. In this analysis, the default scale factor is set to be 1.0. 
Utilizing the signal MC samples, three categories of detector responses are studied with the help of the fast simulation software package, where the detailed information are elaborated below. 

 {\it a. Tracking efficiency} In the fast simulation, two dimensional distributions of transverse momentum $P_{T}$ and polar angle $\cos \theta$ are used to characterize the tracking efficiency, since they are correlated with the level of track bending and hit positions of tracks in the tracker system~\cite{fanyl_stcf}. The reconstruction efficiency for low-momentum tracks ($P_{T}<0.2$ GeV/$c$) will be affected due to stronger electromagnetic multiple scattering, electric field leakage, energy loss, and so on. However, the efficiency for low-momentum tracks can be improved with different techniques in the tracking system design at STCF, or with advanced track finding algorithm.
 
\begin{figure*}[htbp]
\centering
   \includegraphics[width=0.9\textwidth]{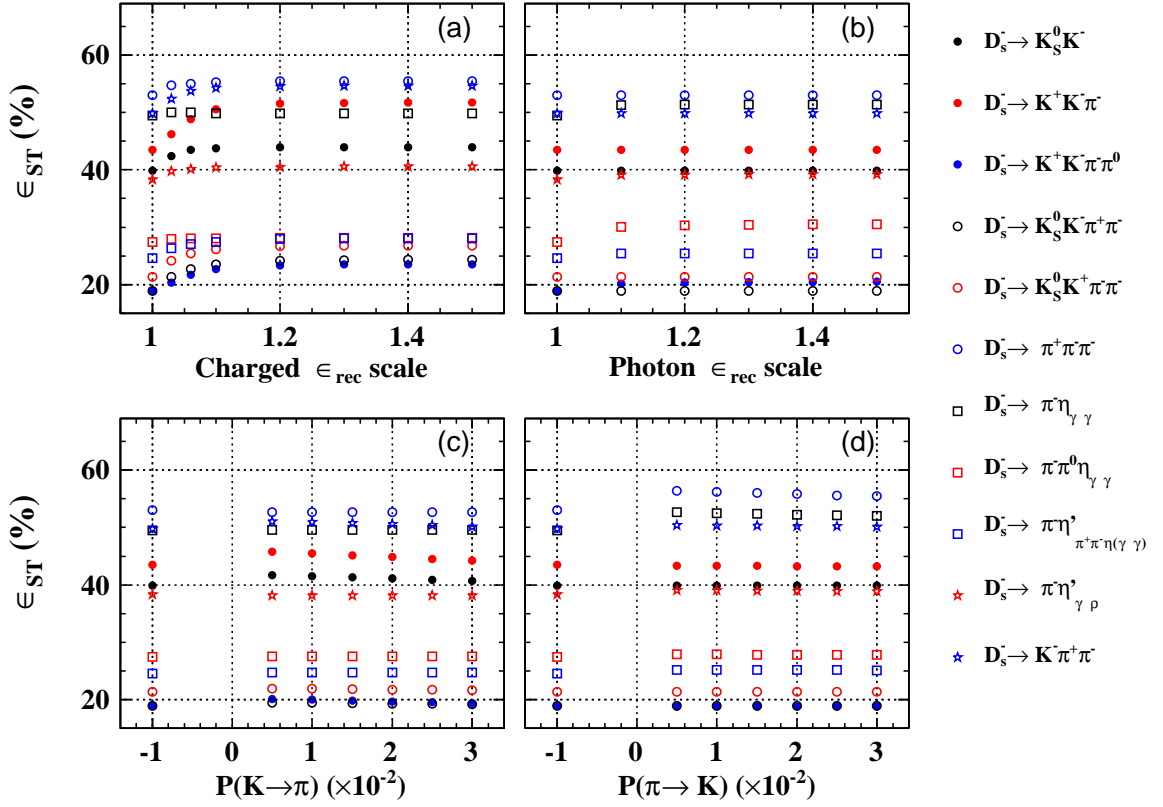}
  \caption{The optimizations of ST efficiencies for the reconstructed efficiencies of (a) charged tracks and (b) photons, and the rates for (c) $K$ misidentified as $\pi$, and (d) $\pi$ misidentified as $K$. The default results are those with the scale factors to be 1.0 in (a) and (b), and the misidentification rates to be -1.0 in (c) and (d).}
\label{opt_track_eff_charged_track}
\end{figure*}

\begin{figure*}[htbp]
\centering
   \includegraphics[width=0.9\textwidth]{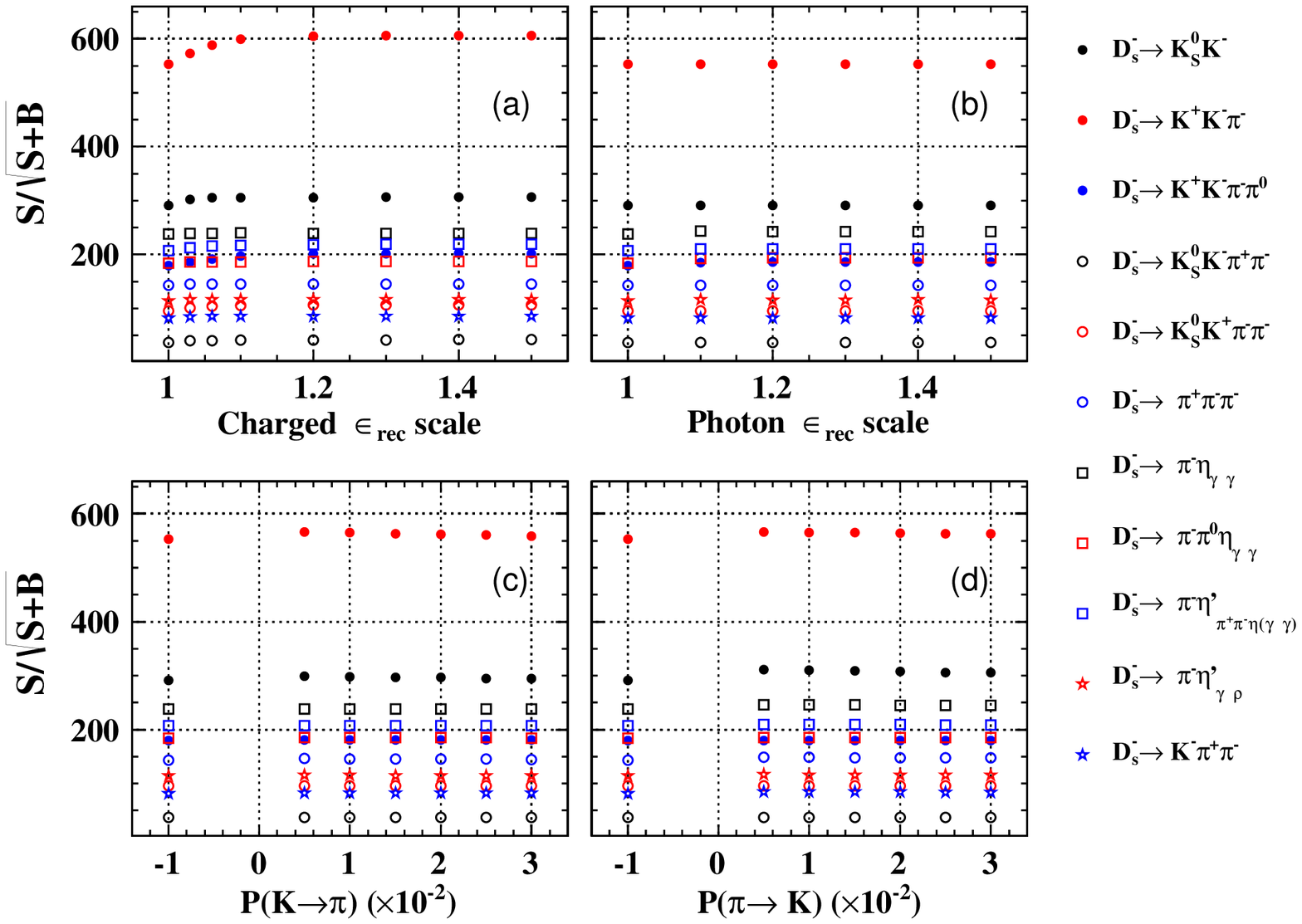}
  \caption{The optimizations of figure-of-merits for the reconstructed efficiencies of (a) charged tracks and (b) photons, and the rates for (c) $K$ misidentified as $\pi$, and (d) $\pi$ misidentified as $K$. The default results are those with the scale factors to be 1.0 in (a) and (b), and the misidentification rates to be -1.0 in (c) and (d).}
\label{opt_track_eff_charged_track_S_B}
\end{figure*}

The tracking efficiency of charged track is varied with a factor from 1.0 to 1.5 in the fast simulation, corresponding to an increasing efficiency up to 50\%. The changes on the ST efficiency and $S/\sqrt{S+B}$ are shown in Figs.~\ref{opt_track_eff_charged_track} (a) and~\ref{opt_track_eff_charged_track_S_B} (a). For high-momentum tracks, the tracking efficiency within acceptance is over 99\%, and this is the reason that this variation has a little influence on the ST efficiency or $S/\sqrt{S+B}$ for $\dstopieta$ mode, where the momentum of the pion is larger than 0.5 GeV/$c$. But for tag modes with low-momentum tracks, the ST efficiency or $S/\sqrt{S+B}$ improves obviously during the variation from 1.0 to 1.1. 

{\it b. Detection efficiency for photon} In the fast simulation, the default detection efficiency for photon is sampled from the performance of the EMC~\cite{fast_simu_tool}. 
Similar to the charged tracks case, a scale factor varied from 1.0 to 1.5 is used to adjust the detection efficiency of photon. 
The related changes on ST efficiency and $S/\sqrt{S+B}$ are shown in Figs.~\ref{opt_track_eff_charged_track} (b) and~\ref{opt_track_eff_charged_track_S_B} (b), respectively. 
It is found that the ST efficiency or $S/\sqrt{S+B}$ can be improved with an optimization factor of 1.1 for the tag modes including photons, such as $\dstopieta$ mode, while there are no effects for the tag modes only with charged tracks. 
The energy and position resolution for photon can also be optimized in the fast simulation. But the default resolutions, 6 mm for position resolution and 2.5\% for energy resolution of a 1 GeV photon~\cite{fast_simu_tool, sanghy_tau_to_kspinu}, can satisfy the physical requirements for our analysis.

{\it c. $\pi/K$ misidentification} At STCF, the $dE/dx$ from the tracking system and the information from the Ring Imaging Cherenkov detector and Detection of Internally Reflected Cherenkov (DIRC) detector in the PID system are used to separate pions and kaons~\cite{fast_simu_tool}. 
The misidentification rate depends on the momentum/direction of the tracks, corresponding to the tracks at 1 GeV/$c$ and with direction perpendicular to the beam. 
The relation of the misidentification rate to the momentum/direction is estimated by {\sc geant4} simulation with the BESIII detector case~\cite{Ablikim:2009aa}. 
We inherit this relationship in the fast simulation, and use it to estimate the misidentification rates for other momenta. 
As the fast simulation provides the function for optimising the $\pi/K$ identification, the $\pi/K$ misidentification rate at 1 GeV/$c$ is varied from 0.5\% to 3.0\%, 
while it is scaled proportionally for other momenta. 
The corresponding ST efficiency is shown in Figs.~\ref{opt_track_eff_charged_track} (c) and (d), while $S/\sqrt{S+B}$ is shown in Figs.~\ref{opt_track_eff_charged_track_S_B} (c) and (d), respectively. 
One can see that the ST efficiencies for $\dstokkpi$ and $\dstopipipi$ mode are obviously increased in Figs.~\ref{opt_track_eff_charged_track} (c) and (d) when the misidentification rates between pions and kaons are decreased, respectively. The optimized misidentification rates of 1.0\% between pion and kaons at 1 GeV/$c$ are expected to fulfill the physical requests for our analysis.

Based on the investigated results discussed above, a set of optimization factors for sub-detector responses is adopted: the reconstructed efficiencies of charged and neutral tracks are improved by 10\%, and the misidentification rate from a $\pi$ ($K$) to $K$ ($\pi$) is set to be 1.0\% at 1 GeV/$c$. Here, the optimization results are consistent with those obtained in the $D_s^+ \to \mu^+ \nu_{\mu}$ case at STCF~\cite{dstotaunu_liujj}. After performing these optimization factors, and following the same procedures in Sect.~\ref{sec_st}, the ST and DT efficiencies are significantly improved, as listed in Table~\ref{com_st_dt_eff}.

\begin{table*}[htbp]
\begin{center}
\caption{Comparisons of ST (DT) efficiencies with the optimization ($\epsilon_{\rm ST(DT)}^{\rm opt}$) or not ($\epsilon_{\rm ST}$), as well as the ST (DT) yields ($N_{\rm ST(DT)}^{\rm opt}$) with the optimization. The intermediated states ($K_S^0$, $\pi^0$, $\eta$, $\eta'$, and $\rho^0$) decays are not included both in the ST and DT efficiencies. All of the uncertainties are statistical only.}
\label{com_st_dt_eff}
\setlength{\extrarowheight}{1.2ex}
 \setlength{\tabcolsep}{7pt}
  \renewcommand{\arraystretch}{1.0}
  \vspace{0.1cm}
\begin{tabular}{l  r@{$\,\pm\,$}l  r@{$\,\pm\,$}l  r@{$\,\pm\,$}l  r@{$\,\pm\,$}l  r@{$\,\pm\,$}l r@{$\,\pm\,$}l }\hline
Mode & \multicolumn{2}{c}{$\epsilon_{\rm ST}$~(\%)} & \multicolumn{2}{c}{$\epsilon_{\rm ST}^{\rm opt}$~(\%)} & \multicolumn{2}{c}{$N_{\rm ST}^{\rm opt}$} & \multicolumn{2}{c}{$\epsilon_{\rm DT}$~(\%)} & \multicolumn{2}{c}{$\epsilon_{\rm DT}^{\rm opt}$~(\%)} & \multicolumn{2}{c}{$N_{\rm DT}^{\rm opt}$}\\\hline
$\dstoksk$	&	40.74 	&	0.30 	&	46.18 	&	0.31 	&	182019 	&	512 	&	26.91	&	0.07	&	30.22	&	0.07	&	1125 	&	48 	\\
$\dstokkpi$	&	42.41 	&	0.16 	&	52.75 	&	0.18 	&	1125600 	&	1567 	&	28.60	&	0.07	&	35.28	&	0.08	&	7467 	&	119 	\\
$\dstokkpipiz$	&	15.80	&	0.17 	&	21.08	&	0.21 	&	567845 	&	2293 	&	11.83	&	0.05	&	16.15	&	0.06	&	3916 	&	102 	\\
$\dstokskpipi$	&	18.04 	&	0.56 	&	21.96 	&	0.64 	&	67306 	&	1151 	&	12.88	&	0.05	&	16.19	&	0.06	&	545 	&	55 	\\
$\dstokskpipim$	&	20.12 	&	0.37 	&	25.93 	&	0.45 	&	110153 	&	802 	&	13.95	&	0.05	&	17.47	&	0.06	&	740 	&	48 	\\
$\dstopipipi$	&	56.09 	&	0.61 	&	61.61 	&	0.63 	&	310194 	&	1284 	&	36.45	&	0.08	&	40.56	&	0.08	&	2086 	&	101 	\\
$\dstopieta$	&	49.30 	&	0.47 	&	55.07 	&	0.50 	&	150116 	&	563 	&	33.95	&	0.07	&	37.79	&	0.08	&	993 	&	42 	\\
$\dstopipizeta$	&	25.65 	&	0.30 	&	27.91 	&	0.31 	&	391979 	&	1748 	&	18.74	&	0.06	&	21.43	&	0.06	&	3037 	&	81 	\\
$\dstopietapgam$	&	24.65 	&	0.29 	&	28.62 	&	0.31 	&	79980 	&	357 	&	16.91	&	0.10	&	19.88	&	0.11	&	528 	&	30 	\\
$\dstopietaprho$	&	33.95 	&	0.49 	&	36.75 	&	0.51 	&	189422 	&	1050 	&	23.44	&	0.09	&	25.81	&	0.09	&	1383 	&	80 	\\
$\dstokpipi$	&	46.12 	&	0.92 	&	54.92 	&	0.97 	&	157161 	&	1132 	&	33.79	&	0.07	&	38.57	&	0.08	&	1123 	&	84 	\\
\hline
\end{tabular}
\vspace{-0.1cm}
\end{center}
\end{table*}

In order to achieve the above optimization parameters, some advanced technologies have been proposed to be the STCF subdetectors, 
such as a thin silicon detector or a micro-pattern gas detector for the inner tracking system, a helium-gas-based cylindrical main drift chamber for the outer tracking system, a Cherenkov based PID system, crystal LYSO or a pure CsI based electromagnetic calorimeter, etc~\cite{shixd_cp}. 
Under the help of the detector geometry management system~\cite{liudong} or {\sc geant4} simulation~\cite{geant4}, each subdetector group can conveniently optimize its subdetector by utilizing the proper material, size, thickness, etc.
For example, the performance for the DIRC-like time of flight (DTOF) detector proposed as the endcap PID detector has been carred out~\cite{dirc_bqi} using {\sc geant4} simulation, and it found that a $\pi/K$ separation power of better than $4\sigma$ at the momentum of 2 GeV/$c$ can be achieved over the entire sensitive area of the DTOF detector, thereby fulfilling the physical requirement of the PID detector at STCF.
In the future, the STCF detector is expected to feature high detection efficiency and resolution, and excellent particle identification capability, and thus meet the requirements of our physical goals.

\section{\boldmath{Results and discussions}}
\label{sec_results}
\subsection{\boldmath{Statistical results}}
Based on the same steps in Sect.~\ref{sec_st}, the ST and DT yields for the optimization case are obtained from the fits to $M_{\rm BC}$ and $\etot$, respectively, as listed in Table~\ref{com_st_dt_eff}. Utilizing Eqs.~(\ref{dtbr}) and (\ref{errdtbr}), the statistical sensitivity for $\mathcal{B}_{D_s^+ \to \tau^+ \nu_{\tau}}$ is determined to be $0.06\times10^{-2}$, which is improved by 14.3\% compared to the case without optimizations. Normalized by a factor of $1/\sqrt{\mathcal{L}}$, the statistical sensitivity is estimated to be $0.02\times10^{-2}$ for the generic MC sample with an expected $\mathcal{L}=$ 1 ab$^{-1}$ collected by STCF per year at $\sqrt{s}=$ 4.009 GeV.

Combining Eq.~(\ref{decayratedstolnu_draft}) with Eq.~(\ref{br_with_decay_width}), and if taking the $|V_{cs}|=0.97320\pm0.00011$ from the CKMfitter group~\cite{pdg} as an input, the $f_{D_s^+}=$ is evaluated to be $(256.3 \pm 0.5_{\rm stat.})$ MeV with the relative statistical uncertainty of 0.2\%. 
Alternatively, using the $f_{D_s^+} = (249.9\pm0.5)$ MeV from LQCD calculations~\cite{fds_flag}, the $|V_{cs}|$ is estimated to be $0.998 \pm 0.003_{\rm stat.}$ with the relative statistical uncertainty of 0.3\%.
These high-precision results will deepen our understanding of strong interactions in the charm sector, and constrain the SM parameters.

\subsection{\boldmath{Systematic uncertainty estimations}}

The systematic uncertainty estimations on the BF of $D_s^+ \to \tau^+ \nu_{\tau}$ are classified into three cases. 

The first one is related with the samples' statistics. According to Ref.~\cite{dsds_crosssetcion}, the ratio of the cross sections of $e^+ e^- \to D_s^+ D^-_s$ at $\sqrt{s}=$ 4.009 GeV over $e^+ e^- \to D_s^{*+} D^-_s$ at $\sqrt{s}=$ 4.178-4.226 GeV is about $0.3$. 
Thus, the statistics for the expected samples with 1 ab$^{-1}$ at $\sqrt{s}=$ 4.009 GeV from STCF will be about a factor of 50 larger than that with 6.32 fb$^{-1}$ at $\sqrt{s}=$ 4.178-4.226 GeV from BESIII~\cite{dstotaunu_hjli}. 
Compared to the systematic uncertainties from the analogous analysis at BESIII~\cite{dstotaunu_hjli}, the corresponding systematic uncertainties based on the control samples can be approximately decreased by a factor of $1/\sqrt{50}$ at STCF if correcting the related difference on the efficiencies between data and MC simulation.  

The systematic uncertainty associated with the signal region of $E_{\rm extra}^{\rm tot}<0.4$ GeV can be studied with the data-MC difference of the acceptant efficiencies obtained from the control sample of the DT events of $D_s^+ \to \pi^+ (\pi^0) \eta$, which is estimated to be 0.11\%. 
With a similar technique, the systematic uncertainty associated with extra charged tracks is thought to be 0.1\%, based on the DT events of $D_s^+ \to \pi^+ \phi (\to K^+ K^-)$ and $D_s^+ \to K^+ \bar{K}^{*}(892)^0(\to K^- \pi^+)$. 
Utilizing the control sample of radiative Bhabha events, the systematic uncertainty from the $e^+$ tracking (PID) efficiency is evaluated to be 0.02\% (0.01)\% after reweighting the corresponding efficiency with the $e^+$ two dimensional (momentum and the polar angle) distribution of signal $D_s^+ \to \tau^+ (e^+ \nu_e \bar{\nu}_{\tau}) \nu_{\tau}$. 
The uncertainty caused by the final state radiation effect is calculated to be 0.03\% with the radiative Bhabha events.

BESIII has measured the BF of $\mathcal{B}_{D_s^+ \to K^0 e^+ \nu_e} = (3.25\pm0.38\pm0.16)\times10^{-3}$~\cite{lilei_dstoklenu} with the data sample of 3.19 fb$^{-1}$ taken at $\sqrt{s}=4.178$ GeV. 
If normalizing its statistical uncertainty and conservatively assuming the related systematic uncertainties to be same, the precision on $\mathcal{B}_{D_s^+ \to K^0 e^+ \nu_e}$ at STCF can reach to 4.8\%. 
The size of the $D_s^+ \to K^0_L e^+ \nu_e$ background can be examined by sampling $\mathcal{B}_{D_s^+ \to K^0_L e^+ \nu_e}$ $10^{4}$ times with a random Gaussian function based on its uncertainty. The distribution of the relative difference on the DT yield can be fitted by a Gaussian function, and the width is taken as the systematic uncertainty from the size of the $D_s^+ \to K^0_L e^+ \nu_e$ background, which is set to be 0.5\%. 
Since it is the dominant systematic uncertainty, the improved precision on $\mathcal{B}_{D_s^+ \to K^0 e^+ \nu_e}$ at STCF in the future is necessary for our analysis. Compared to the previous measurement of $\mathcal{B}_{D_s^+ \to K^0_L e^+ \nu_e}$~\cite{lilei_dstoklenu}, one can choose control samples at STCF to study the systematic uncertainties from the requirements of the largest energy of any unused photon and the invariant mass of $K^0 e^+$, instead of varying the corresponding range. Based on the control samples with larger sizes at STCF, it can also be improved for the systematic uncertainties from the $e^+$ tracking and PID efficiencies and $K^0$ reconstruction efficiencies.

The systematic uncertainty from the limited MC statistics is estimated to be 0.27\% based on a 1 ab$^{-1}$ MC sample at STCF. The uncertainty from the background fluctuation of the fitted ST yield is assigned to be 0.07\%. The size of the $D^- \to K_S^0 \pi^-$ background in the $D_s^- \to K_S^0 K^-$ tag can be varied by $10^4$ times based on a Gaussian function with its uncertainty, and the width of relative difference on the DT yield is assigned as the systematic uncertainty from the size of the $D^- \to K_S^0 \pi^-$ background in the $D_s^- \to K_S^0 K^-$ tag mode, which is evaluated to be 0.01\%. 

The second one is the rest parts of the systematic uncertainties in the selection, mainly from the fit procedure~\cite{dstotaunu_hjli}. 
The systematic uncertainty in the ST yield consists of the fit range, the signal and background shapes, and the bin size. 
The fit range can be altered by $1\sigma$ of $M_{\rm ST}$ reduced from both sides of the nominal range. The signal shapes obtained from the generic MC sample can be replaced with those from the signal MC sample. The background shape can be changed to a different order of the Chebychev function. The bin size can be doubled or halved. The difference in the ratio of the ST yield over the ST efficiency for each variation in a given ST mode is added in quadrature, and then weighted by the ST yields, which is assigned to be 0.37\%. 
 
Due to different charge and neutral multiplicities, the ST efficiencies estimated with the generic and signal MC samples are expected to be different slightly. Thus, the uncertainty associated with the ST efficiency can not be fully cancelled, which results in a so called ``tag bias" uncertainty.
The tracking and PID efficiencies in different multiplicities at STCF are conservatively assumed to be same as those at BESIII~\cite{dstotaunu_hajime}. And then combining the difference on the ST efficiency estimated with the above two MC samples, the uncertainty from the tag bias is set to be 0.26\%.

The alternative shape of $D_s^+ \to X e^+ \nu_e$ background can be chosen by varying the proportions of the six main components, through sampling the corresponding BF~\cite{lilei_dstoklenu, etaenu_1, etaenu_2, phienu_3, phienu_2, phienu_4, f0enu_1, alex_1128_2018_BESIII_Collaboration} $10^4$ times based on a Gaussian function given by its uncertainty. Here, the MC simulation study shows that the six main components are $D_s^+ \to \eta e^+ \nu_e$, $\eta' e^+ \nu_e$, $\phi e^+ \nu_e$, $f_0(980) e^+ \nu_e$, $K^*(892)^0 e^+ \nu_e$ and $K_S^0 e^+ \nu_e$ decays. The width of the relative difference of the ST efficiency weighted by the ST yields is taken as the systematic uncertainty from the fixed $D_s^+ \to X e^+ \nu_e$ background shape, which is 0.1\%.

The systematic uncertainty from the non-$D_s^-$ background can be estimated with an alternative shape that obtained from the background events in the $M_{\rm ST}$ signal region from the generic MC sample, which is set to be 0.07\%. 
The uncertainty from the quoted BF of $\tau^+ \to e^+ \nu_e \bar{\nu}_{\tau}$ is 0.2\%\cite{pdg}.

The third one is that the current experiments~\cite{pdg, dstotaunu_hjli} has not considered. 
In the $D_s^+ \to \tau^+ \nu_{\tau}$ decay, the $\tau^+$ lepton only has a kinetic energy of 9.3 MeV in the $D_s^+$ rest frame, so the effect from the radiative $D_s^+ \to \gamma \tau^+ \nu_{\tau}$ decay has been negligible in the current experiments~\cite{pdg, dstotaunu_hjli}. 
But the relative statistical precision of $\mathcal{B}_{D_s^+ \to \tau^+ \nu_{\tau}}$ at STCF is up to 0.4\%, the effect from the radiative $D_s^+ \to \gamma \tau^+ \nu_{\tau}$ decay should be considered. 
In the future, we can perform the analysis of $D_s^+ \to \gamma \tau^+ \nu_{\tau}$ with a 0.01 GeV cutoff on the radiative photon energy at STCF, 
of which the statistical uncertainty can be conservatively estimated to be 10\% if assuming its BF is two magnitudes lower than $D_s^+ \to \tau^+ \nu_{\tau}$. 
Since the radiative fraction from the theoretical calculations is no more than 0.8\% with the minimum energy of photon to be 0.01 GeV~\cite{dstogamtaunu1, dstogamtaunu2, dstogamtaunu3, dstogamtaunu4}, the radiative effect in the nominal analysis is evaluated to be less than 0.1\%. 

The total systematic uncertainty on the $\mathcal{B}_{D_s^+ \to \tau^+ \nu_{\tau}}$ is conservatively estimated to be 1.0\%.

According to Eqs.~(\ref{decayratedstolnu_draft}) and (\ref{br_with_decay_width}), the relative systematic uncertainty on $f_{D_s^+}$, $\frac{\Delta f_{D_s^+}}{f_{D_s^+}}$ is estimated to be 0.6\% with the formula,
\begin{equation}
\frac{\Delta f_{D_s^+}}{f_{D_s^+}} = \sqrt{\left(\frac{1}{2}\frac{\Delta \tau_{D_s^+}}{\tau_{D_s^+}}\right)^2 + \left(\frac{1}{2}\frac{\Delta \mathcal{B}}{\mathcal{B}}\right)^2 + \left(\frac{\Delta |V_{cs}|}{|V_{cs}|}\right)^2}, 
\end{equation}
where $\tau_{D_s^+}=(504\pm4) \times 10^{-15}$ s~\cite{pdg}, the relative systematic uncertainty on $\mathcal{B}_{D_s^+ \to \tau^+ \nu_\tau}$ is $\frac{\Delta \mathcal{B}}{\mathcal{B}}=1.0\%$, and the input value $|V_{cs}|=0.97320\pm 0.00011$ is from the CKMfitter group~\cite{pdg}. 
Similarly, the relative systematic uncertainty on $|V_{cs}|$, $\frac{\Delta |V_{cs}|}{|V_{cs}|}$ is estimated to be 0.7\% with the formula,
\begin{equation}
\frac{\Delta |V_{cs}|}{|V_{cs}|} = \sqrt{\left(\frac{1}{2}\frac{\Delta \tau_{D_s^+}}{\tau_{D_s^+}}\right)^2 + \left(\frac{1}{2}\frac{\Delta \mathcal{B}}{\mathcal{B}}\right)^2 + \left(\frac{\Delta f_{D_s^+}}{f_{D_s^+}}\right)^2}, 
\end{equation}
where the input value $f_{D_s^+}=(249.9\pm0.5)$ MeV is from LQCD calculations~\cite{fds_flag}.

\subsection{\boldmath{Combined results at STCF}}

Combining the prospect of $\mathcal{B}_{D_s^+ \to \mu^+ \nu_{\mu}}$ at STCF in Ref.~\cite{dstotaunu_liujj}, the ratio of $\mathcal{B}_{D_s^+ \to \tau^+ \nu_{\tau}}$ over $\mathcal{B}_{D_s^+ \to \mu^+ \nu_{\mu}}$ is calculated to be $9.79\pm0.05_{\rm stat.}\pm0.11_{\rm syst.}$, 
where the common systematic uncertainties of the two decay modes are assumed to be cancelled out, such as those from the limited MC statistics, the fit range, the signal and background shapes, the bin size and the background fluctuation in the ST yield, and the tag bias.
Compared to the current experimental value of $9.67\pm0.34$~\cite{shenxy_talk_charm2020}, the precision of the ratio can be improved by a factor of 2.9 at STCF, which will provide the most stringent test of the $\tau$-$\mu$ LFU in heavy quark decays~\cite{lihb_charm_review, hflav_group}.

The expected values of $f_{D_s^+}$ and $|V_{cs}|$ at STCF are summarized in Table~\ref{average_results_stcf} from $D_s^+ \to \tau^+ \nu_{\tau}$ in this analysis and $D_s^+ \to \mu^+ \nu_{\mu}$ in Ref.~\cite{dstotaunu_liujj}. The averaged results of inverse uncertainty weighted values are estimated to be 
$f_{D_s^+} =  (256.0\pm0.3_{\rm stat.}\pm1.5_{\rm syst.})$ MeV and
$|V_{cs}|  = 0.997\pm0.002_{\rm stat.}\pm0.006_{\rm syst.}$, where the mean values are evaluated by weighting both statistical and uncorrelated systematic uncertainties. Here, the systematic uncertainties between two decay modes are assumed to be uncorrelated, excluding the above common ones and those from the external input values. One can see that the statistical precision of $f_{D_s^+}$ at STCF can be improved by a factor of 1.7 compared to the current LQCD calculations of $f_{D_s^+}=(249.9\pm0.5)$ MeV~\cite{fds_flag}.
It should be noted that currently, only rough and conservative estimations of systematic uncertainties are available for $D_s^+ \to \tau^+ \nu_{\tau}$ in this analysis and $D_s^+ \to \mu^+ \nu_{\mu}$ in Ref.~\cite{dstotaunu_liujj}. Therefore, the systematic uncertainties presented in this analysis will be optimized further while the design of the STCF detector is completed. In addition, the external input value of the $\tau_{D_s^+}$ introduces a relative uncertainty of 0.4\%, which is the dominant systematic uncertainty in the averaged results of $f_{D_s^+}$ and $|V_{cs}|$. It will be much helpful for our precision at STCF if the uncertainty of the $\tau_{D_s^+}$ from LHCb or other experiments can be improved in the future.

\begin{table}[htbp]
\centering
\caption{The expected values of $f_{D_s^+}$ and $|V_{cs}|$ at STCF from $D_s^+ \to \tau^+ \nu_{\tau}$ in this analysis and $D_s^+ \to \mu^+ \nu_{\mu}$ in Ref.~\cite{dstotaunu_liujj} at STCF. The first and second uncertainties are the statistical and systematic uncertainties, respectively. The dominant systematic uncertainty arises from the external input of the $D_s^+$ lifetime. ``Average" denotes the averaged result of inverse uncertainty weighted value, where the mean value is estimated by weighting both statistical and uncorrelated systematic uncertainties. The detailed information is described in the text.}
\label{average_results_stcf}
 \setlength{\extrarowheight}{1.2ex}
 \setlength{\tabcolsep}{7pt}
 \renewcommand{\arraystretch}{1.0}
  \vspace{0.1cm}
\begin{tabular}{l c c}\hline
Signal mode &  $f_{D_s^+}$ (MeV) & $|V_{cs}|$ \\\hline
$D_s^+ \to \tau^+ \nu_{\tau}$ &  $256.3\pm0.5\pm1.5$ & $0.998\pm0.003\pm0.007$ \\ 
$D_s^+ \to \mu^+ \nu_{\mu}$   &  $255.8\pm0.4\pm1.5$ & $0.996\pm0.002\pm0.007$ \\\hline 
Average                       &  $256.0\pm0.3\pm1.5$ & $0.997\pm0.002\pm0.006$ \\ 
\hline
\end{tabular}
 \vspace{-0.1cm}
\end{table}
 
\subsection{\boldmath{Comparisons with other sensitive results}}
Table~\ref{com_results_stcf_others} lists the comparisons of the relative expected precision on the measurements of $\mathcal{B}_{D_s^+ \to \tau^+ \nu_{\tau}}$, $f_{D_s^+}$, $|V_{cs}|$, and the ratio of $\mathcal{B}_{D_s^+ \to \tau^+ \nu_{\tau}}$ over $\mathcal{B}_{D_s^+ \to \mu^+ \nu_{\mu}}$ among BESIII with an expected 6 fb$^{-1}$ at $\sqrt{s}=$ 4.178 GeV~\cite{bes3_future}, BelleII with an expected 50 ab$^{-1}$ at $\Upsilon(nS)$~\cite{bes3_future}, and this work with an expected 1 ab$^{-1}$ at STCF. 
One can see that the results at STCF with 1 ab$^{-1}$ will play a decisive role in the future world average values. 
 
\begin{table*}[htbp]
\begin{center}
\caption{Comparisons of the relative expected precision on the measurements of $\mathcal{B}_{D_s^+ \to \tau^+ \nu_{\tau}}$, $f_{D_s^+}$, $|V_{cs}|$, and the ratio of $\mathcal{B}_{D_s^+\to\tau^+ \nu_{\tau}}$ over $\mathcal{B}_{D_s^+\to\mu^+ \nu_{\mu}}$ among BESIII~\cite{bes3_future}, BelleII~\cite{bes3_future}, and STCF experiments. The relative statistical and systematic uncertainties for $\mathcal{B}_{D_s^+ \to \mu^+ \nu_{\mu}}$ are estimated to be 0.3\% and 1.0\% at STCF~\cite{dstotaunu_liujj}. (Here, ``-" indicates not available).}
\label{com_results_stcf_others}
\setlength{\extrarowheight}{1.2ex}
 \setlength{\tabcolsep}{7pt}
  \renewcommand{\arraystretch}{1.6}
  \vspace{0.1cm}
\begin{tabular}{l |c c| c c| c c }\hline
\multirow{2}{*}{Source}	&	\multicolumn{2}{c|}{BESIII~\cite{bes3_future}}			&	\multicolumn{2}{c|}{BelleII~\cite{bes3_future}}			&	\multicolumn{2}{c}{This work at STCF}			\\\cline{2-7}
	&	\multicolumn{2}{c|}{6 fb$^{-1}$ at 4.178 GeV}			&	\multicolumn{2}{c|}{50 ab$^{-1}$ at $\Upsilon(nS)$}			&	\multicolumn{2}{c}{1 ab$^{-1}$ at 4.009 GeV}			\\\hline
$\mathcal{B}_{D_s^+ \to \tau^+ \nu_{\tau}}$	&	1.6\%$_{\rm stat.}$	&	2.4\%$_{\rm syst.}$	&	$0.6\%_{\rm stat.}$	&	$2.7\%_{\rm syst.}$	&	$0.3\%_{\rm stat.}$	&	$1.0\%_{\rm syst.}$	\\
$f_{D_s^+}$ (MeV)	&	$0.9\%_{\rm stat.}$	&	$1.4\%_{\rm syst.}$	&	$-$	&	$-$	&	$0.2\%_{\rm stat.}$	&	$0.6\%_{\rm syst.}$	\\
$|V_{cs}|$	&	$0.9\%_{\rm stat.}$	&	$1.4\%_{\rm syst.}$	&	$-$	&	$-$	&	$0.3\%_{\rm stat.}$	&	$0.7\%_{\rm syst.}$	\\
$\frac{\mathcal{B}_{D_s^+\to\tau^+ \nu_{\tau}}}{\mathcal{B}_{D_s^+\to\mu^+ \nu_{\mu}}}$	&	$2.6\%_{\rm stat.}$	&	$2.8\%_{\rm syst.}$	&	$0.9\%_{\rm stat.}$	&	$3.2\%_{\rm syst.}$	&	$0.5\%_{\rm stat.}$	&	$1.1\%_{\rm syst.}$	\\
\hline
\end{tabular}
\vspace{-0.1cm}
\end{center}
\end{table*}

\section{\boldmath{Summary}}
\label{sec_summary}
In brief, based on an expected generic MC sample of 1 ab$^{-1}$ at $\sqrt{s}=$4.009 GeV at STCF, the statistical sensitivity of the absolute BF of $D_s^+ \to \tau^+ \nu_{\tau}$ is determined to be $2\times10^{-4}$ with the optimization factors for sub-detector responses by means of the fast simulation. 
Combined with the results of $\mathcal{B}_{D_s^+ \to \mu^+ \nu_{\mu}}$ at STCF, the relative statistical sensitivity of the LFU can reach at a level of 0.5\%. 
The decay constant $f_{D_s^+}$ and CKM matrix element $|V_{cs}|$ are also extracted separately. These results are important to calibrate LQCD calculations of $f_{D_s^+}$, test the unitarity of the CKM matrix and LFU in $\tau$-$\mu$ flavors with higher precision.

.

\section*{\boldmath Acknowledgments}

The authors thank the supercomputing center of USTC and Hefei Comprehensive National Science Center of their strong support. 
This work is supported by National Natural Science Foundation of China (NSFC) under Contracts No. 12105077, No. 12122509, No. 11805037, No. 11875122, No. 11625523; Joint Large-Scale Scientific Facility Funds of the NSFC under Contracts No. U1832121; The Double First-Class university project foundation of USTC; USTC Research Funds of the Double First-Class Initiative No. YD2030002005; Excellent Youth Foundation of Henan Province No. 212300410010; The youth talent support program of Henan Province No. ZYQR201912178; The Program for Innovative Research Team in University of Henan Province No. 19IRTSTHN018; International partnership program of the Chinese Academy of Sciences Grant No. 211134KYSB20200057.


\end{sloppypar}
\end{document}